\documentclass[prd,showpacs,aps,preprint]{revtex4}
\usepackage{graphicx}
\usepackage{dcolumn}
\usepackage{amsfonts}
\usepackage{amsmath}

\newcommand{\ep}{\epsilon}
\newcommand{\vep}{\varepsilon}

\newcommand{\lraw}{\longrightarrow}
\newcommand{\llaw}{\longleftarrow}
\newcommand{\pa}{\partial}
\newcommand{\td}{\tilde}

\newcommand{\beq}[1]{\begin{eqnarray}\label{#1}}
\newcommand{\eeq}{\end{eqnarray}}

\newcommand{\llraw}{-\!-\!\!\!\lraw}
\newcommand{\lllaw}{\llaw\!\!\!-\!-}
\newcommand{\CN}[1]{{\cal N}=#1}
\newcommand{\nn}{\nonumber}

\begin{document}
\title{A note on brane/flux annihilation and dS vacua in string theory}
\author{Xiao-Jun Wang}
\email{wangxj@ustc.edu.cn}
\affiliation{\centerline{Interdisciplinary Center for Theoretical
Study} \centerline{University of Science and Technology of China}
\centerline{AnHui, HeFei 230026, China}}
\author{Neng-Chao Xiao}
\affiliation{\centerline{Department of Modern Physics}
\centerline{University of Science and Technology of China}
\centerline{AnHui, HeFei 230026, China}}

\begin{abstract}
We reconsider the dynamics of $p$ anti-D3 branes inside the
Klebanov-Strassler geometry, in which $M$ units of R-R 3-form flux
and $K$ units of NS-NS 3-form flux are presented in deformed
conifold. We find that anti-D3 branes blow up into a spherical
D5-brane at weak string coupling via quantum tunnelling. The
D5-brane can be either stable or unstable, depending on number of
background flux. The nucleation rate of D5-brane is suppressed by
$\exp{\{-Mp^2\}}$. The classical mechanically the evolution of
unstable D5-brane annihilates one unit of R-R flux and ends with
$(K-p)$ D3-branes. This observation is consistent with one by
Kachru, Pearson and Verlinde, who shew that anti-D3 branes in KS
geometry can blow up into a spherical NS5 brane at strong string
coupling, because NS5-brane is lighter that D5-brane at strong
string coupling. We also argue that the system can end with a
meta-stable dS vacuum by fine tuning of number of background flux.
\end{abstract}
 \preprint{USTC-ICTS-04-21} \maketitle

\section{Introduction}

There has recently been a great deal of interest on the KKLT
model\cite{KKLT03} since this model provides an explicit realization on de
Sitter (dS) vacua in string theory and is consistent with stabilization of
various moduli (in particular, including volume modulus). KKLT, following
earlier works on flux compactification in IIB
theory\cite{BP00,FMSW00,GKP02}, noted that the volume modulus can be fixed
by considering some non-perturbative effects (e.g., the gluino
condensation). While the flux compactification yields supersymmetric AdS
vacua in $\CN{1}$ supergravity. This AdS minimum may be uplifted to
non-supersymmetric dS one by introducing anti-D3 branes into
Klebanove-Strassler (KS) geometry\cite{KS00}, which is warped background
described by $M$ units of R-R 3-form flux and $K$ units of NS-NS 3-form
flux presented in the deformed conifold. The key point is that there is
extra energy contribution from tension of $\overline{\rm D3}$-branes and
extra fluxes. Moreover, the dynamics of $\overline{\rm D3}$-branes is also
important because the potential of world-volume scalar may in general
contribute a negative energy.

The dynamics of $p$ $\overline{\rm D3}$-branes ($p\ll K,M$) in KS
geometry has been studied by Kachru, Pearson and Verlinde (KPV)
\cite{KPV01} in strong string coupling region. A small negative
minimum of world-volume scalar potential was shown to correspond
to a non-supersymmetric NS5-brane ``giant graviton''
configuration. In other words, the $\overline{\rm D3}$-branes are
expanded into a spherical NS5-brane due to the presence of
background NS-NS flux, according to the effect first observed by
Myers\cite{Myers99}. The NS5-brane is quantum mechanically
unstable via tunnelling to a nearby supersymmetric vacuum. The
decay rate is exponentially suppressed. Consequently the
meta-stable dS vacuum has longer life and world-volume scalar
potential only sightly corrects dS minimum in this regime.

The purpose of this present note is to address natural questions
that what happens in weak string coupling region, and what role
the R-R flux plays. We expect that $\overline{\rm D3}$-branes may
be expanded into a non-supersymmetric D5-brane ``giant graviton''
configuration by R-R flux, as similar as NS5-brane by NS-NS flux.
The D5-brane should be nucleated at weak string coupling because
in this region it is lighter than NS5 and is easier to be formed.
At the first glance we should worry whether the Myers' effect is
induced by R-R flux since $\overline{\rm D3}$-branes in KS
background are quickly driven to the end of the throat (or the
apex of the deformed conifold) \cite{KPV01}, where $\star F_{(3)}$
disappears. We notice, however, that the apex of the conifold is
fuzzy due to non-Abelian effect of world-volume scalar on
$\overline{\rm D3}$-branes. In other words, we can not say where
is exact location of the apex. In this sense the Myers' effect is
indeed induced by R-R flux. Consequently $\overline{\rm
D3}$-branes blow up into a spherical D5-brane.

We shall show that the dynamical process to nucleate spherical
D5-brane is in essential different from one to form NS5-brane. It
is classically impossible to nucleate D5-brane from $\overline{\rm
D3}$-branes without extra initial conditions input (e.g., opposite
motion among of $\overline{\rm D3}$-branes). Quantum mechanically,
however, $\overline{\rm D3}$-branes can tunnel to a D5-brane
``giant graviton'' configuration. The tunnelling rate is
exponentially suppressed if the number of $\overline{\rm
D3}$-branes is larger. Furthermore, D5-brane after quantum brith
can be classically either stable or unstable, depending on the
numbers of background fluxes. The evolution of unstable D5-brane
will annihilate one unit of R-R flux and end with $(K-p)$
D3-branes. Geometrically, the KS background removes the
singularity of the conifold by blowing up a three-sphere at the
apex of the conifold$^1$.
\footnotetext{The conifold can be described as a cone over the
space $T^{1,1}$, which can be topologically thought of as
$S^3\times S^2$.} The KPV's NS5-brane is further nucleated by
distributing $\overline{\rm D3}$-branes on a two-sphere inside
that $S^3$. The spherical D5-brane, by contrast, is nucleated by
distributing $\overline{\rm D3}$-branes on $S^2$ transverse to the
$S^3$.

This note is organized as follows: In section 2, We give a brief
review on embedding KS geometry into F-theory compactification.
The dynamics of a few of $\overline{\rm D3}$-branes probing in KS
background is considered in section 3. In section 4 we compute
rate of the quantum tunnelling to nucleate a D5-brane ``giant
graviton'' configuration from $\overline{\rm D3}$-branes. Section
5 lists a summary and some discussions, including the discussion
on dS vacua in our model.

\section{Embedding Klebanove-Strassler geometry into F-theory
compactification}

Klebanov-Strassler(KS) background\cite{KS00} is vacuum solution
IIB supersgravity. It is produced by placing $N$ D3-branes and $M$
D5-branes wrapping on a collapsing supersymmetric two-cycle at the
apex of the deformed conifold defined by
\beq{1}\sum_{i=1}^4z_i^2=\vep^2,\eeq
in which the singularity of the conifold is removed through the
blowing-up of the $S^3$ of $T^{1,1}$. Here $\vep$ controls the
size of the $S^3$ and has dimensions of length $3/2$. At the end
of evolution the D3-branes are dissolved and their charge is
carried by R-R and NS-NS 3-form fluxes.  Because $\overline{\rm
D3}$-branes will be quickly driven to the end of the throat when
they are put into KS geometry\cite{KPV01}, we only focus on the
geometry near the apex of KS background$^1$ \cite{HKO01}
\footnotetext[1]{Our expressions on
metric and various fluxes are different from one in refs.
\cite{KS00,HKO01} by a rescale of coordinates, $x^a\to
2^{1/6}x^a,\;\; y_i\to 48^{1/6}y_i$.}:
\beq{2}ds^2&\stackrel{\tau\to
0}{\lraw}&\frac{\vep^{4/3}}{\sqrt{(a_0-a_1\tau^2+a_2\tau^4)}\;
g_sMl_s^2} dx_adx_a+\sqrt{a_0-a_1\tau^2+a_2\tau^4}\;g_sMl_s^2
\left\{d\tau^2
+(g^5)^2\right.\nn \\
&&\quad\quad\left.+2(g^3)^2+2(g^4)^2
+\frac{1}{2}\tau^2[(g^1)^2+(g^2)^2]\right\}, \nn \\
F_{(3)}&\stackrel{\tau\to 0}{\lraw}& 2\sqrt{3}\;Ml_s^2\{g^5\wedge
g^3\wedge g^4 +\frac{\tau^2}{12}g^5\wedge g^1\wedge g^2
+\frac{\tau}{6}g^5\wedge (g^1\wedge g^3+g^2\wedge g^4)\},\nn \\
H_{(3)}&\stackrel{\tau\to 0}{\lraw}& 2\sqrt{3}\;g_sMl_s^2\{d\tau
\wedge(\frac{1}{3}g^3\wedge g^4 +\frac{\tau^2}{4}g^1\wedge g^2)
+\frac{\tau}{6}g^5\wedge (g^1\wedge g^3+g^2\wedge g^4)\}, \\
{\cal F}_{(5)}&=&B_{(2)}\wedge F_{(3)} \;\stackrel{\tau\to
0}{\lraw}\;\frac{4g_sM^2l_s^4}{3}\tau^3g^1\wedge g^2\wedge
g^3\wedge g^4\wedge g^5, \nn \\
\star F_{(3)}&\stackrel{\tau\to 0}{\lraw}&
\frac{2\sqrt{3}\vep^{8/3}}{a_0g_s^2Ml_s^2}dx^0\wedge dx^1\wedge
dx^2\wedge dx^2\wedge \{d\tau\wedge(\frac{1}{3}g^3\wedge
g^4+\frac{\tau^2}{4} g^1\wedge g^2) \nn
\\&&\hspace{1in}+\frac{\tau}{6}g^5\wedge (g^1\wedge g^3+g^2\wedge
g^4)\}, \nn
\eeq
together with constant dilation field, where the constants
$a_0\simeq 0.718$, $a_1=3^{-4/3}$ and $a_2=2^{1/3}/3^{7/3}$,
$g^i\;(i=1,...,5)$ are one-form basis on $T^{1,1}$\cite{MT99},
\beq{3}
 g^1&=&-\frac{1}{\sqrt{2}}(\sin{\theta_1}d\phi_1+\cos{\psi}
   \sin{\theta_2}d\phi_2-\sin{\psi}d\theta_2), \nonumber \\
 g^2&=&\frac{1}{\sqrt{2}}(d\theta_1-\sin{\psi}\sin{\theta_2}
   d\phi_2-\cos{\psi}d\theta_2), \nonumber \\
 g^3&=&-\frac{1}{\sqrt{2}}(\sin{\theta_1}d\phi_1-\cos{\psi}
   \sin{\theta_2}d\phi_2+\sin{\psi}d\theta_2), \nonumber \\
 g^4&=&\frac{1}{\sqrt{2}}(d\theta_1+\sin{\psi}\sin{\theta_2}
   d\phi_2+\cos{\psi}d\theta_2), \nonumber \\
 g^5&=&d\psi+\cos{\theta_1}d\phi_1+\cos{\theta_2}d\phi_2.
\eeq
The transverse geometry near the apex of deformed conifold is thus
the $S^2$ fibered over the $S^3$. In the above set of coordinates
the two- and three-cycles are respectively parameterized by
\beq{4}
&S^2&:\quad\phi_1=-\phi_2,\quad\quad
\theta_1=(-1)^n\theta_2,\quad\quad\psi=(2n+1)\pi, \nn \\
&S^3&:\quad\theta_1=\phi_1=0,
\eeq
with $n$ integer. It is easy to check that the two-cycle defined
by the above parameterization is that of minimal volume. It means
D5-brane wraps on this cycle is with minimal energy. Hence it is a
supersymmetric two-cycle\cite{VLM02}.

In order to embedding KS geometry into F-theory compactification,
one has to incorporate the tadpole cancellation condition
\beq{5}\frac{\chi(X)}{24}=N_{D3}-N_{\overline{D3}}
+\frac{1}{2\kappa_{10}^2T_3}\int_{\cal M}H_{(3)}\wedge F_{(3)},
\eeq
where $T_3$ is the D3-brane tension, and $\chi(X)$ is the Euler
number of the CY fourfold $X$ that specifies the F-theory
compactification.

Dirac quantization implies that those fluxes, integrated over all
of the three-cycles of the CY, be integers. From Eq.~(\ref{2}) one
see that the KS solution corresponds to the placing of $M$ units
of R-R flux through the $A$-cycle spanned by $g^3,\;g^4,\;g^5$,
and $K$ units of NS-NS flux through the dual $B$-cycle$^2$ spanned
by $g^1,\;g^2,\;\tau$,
\footnotetext[2]{In this type of scheme of compactification, KS
geometry is cut and glued smoothly with a compacted
manifold\cite{Verlinde00}. Hence radius coordinate $\tau$ varies
in finite region.}
\beq{6}\frac{1}{4\pi^2}\int_A F_{(3)}&=&M, \nn \\
\frac{1}{4\pi^2}\int_B H_{(3)}&=&-K.
\eeq
In terms of choosing $M$ and $K$ to let $MK=\chi/24$, D3-brane
charge is conserved without extra D3-brane inserted.

\section{Dynamics of the probe $\overline{D3}$-branes}

Let us assume that there is little mismatch between Euler number
of CY fourfold and the numbers of fluxes, e.g., $MK-p=\chi/24$.
Then two side of tadpole condition~(\ref{6}) has to be balanced by
introduce $p$ anti-D3 branes, which is driven to the end of throat
of KS geometry. The characteristic size of the geometry is order
$\sqrt{g_sM}l_s$, while the backreaction from the $p$
$\overline{D3}$-branes can be estimated to extend over a region of
order $\sqrt{g_sp}l_s$. Hence the distortion of KS geometry due to
presence of the $\overline{D3}$-branes can be ignored as long as
$p\ll M$\cite{KPV01}.

\subsection{World-volume action}
The low energy dynamics of $N$ coincident anti-Dp branes is
described by the following non-Abelian DBI
action\cite{Myers99,TR99}
\beq{7}
S_{BI}=-T_p\int d^{p+1}\xi\; {\rm STr}\left(e^{-\phi}\sqrt{-{\rm
det} (P[E_{ab}+E_{ai}(Q^{-1}-\delta)^{ij}E_{jb}]+2\pi l_s^2
F_{ab}) {\rm det}(Q^i_{\ j})}\right),
\eeq
with
\beq{8}E_{\mu\nu}=G_{\mu\nu}+B_{\mu\nu},\hspace{1in}
Q^i_{\ j}=\delta^i_{\ j}+2\pi il_s^2[\Phi^i,\Phi^k]E_{kj},
\eeq
plus corresponding Chern-Simons action
\beq{9}S_{CS}=-\mu_p\int {\rm STr}\left(P[e^{2\pi il_s^2
\rm{i}_{_\Phi}\rm{i}_{_\Phi}}(\sum_n C^{(n)}e^B)]e^{2\pi l_s^2
F}\right)
\eeq
and supplemented by proper fermionic part. Here
$G_{\mu\nu},\;B_{\mu\nu}$ and $C^{(n)}$ are background metric,
NS-NS 2-form and R-R $n$-form potential respectively. $P[\cdots]$
denotes the pullback of the enclosed spacetime tensors to the
worldsheet of D-strings. The transverse scalars $\Phi^i$ are
$N\times N$ matrices in the adjoint representation of the $U(N)$
worldsheet gauge symmetry. STr($\cdots$) denotes a symmetrical
trace in $U(N)$ gauge group. Finally, the operator i$_{_\Phi}$ is
defined by
\beq{10}
{\rm i}_{_\Phi}{\rm i}_{_\Phi}C^{(n)}=\frac{1}{(n-2)!}\Phi^{i_1}
\Phi^{i_2}C^{(n)}_{i_1i_2i_3\cdots i_n}dx^{i_3}\wedge\cdots\wedge
dx^{i_n}.
\eeq

We consider a two-sphere solution which just corresponds to the
$S^2$ defined by Eq.~(\ref{4}). It is not hard to see that the
$S^2$ with $n=0$ is equivalent to impose the condition
$g^3=g^4=g^5=0$. In language of matrix model from BI-action, it
means to reduce six transverse scalar $\Phi^i$ to three by
imposing certain conditions. This is always possible in search of
the static solutions. With the above conditions, the transverse
metric, $\star F_{(3)}$ and $H_{(3)}$ on two-cycle are reduced by
\beq{11}&&d\tau^2+\frac{1}{2}\tau^2[(g^1)^2+(g^2)^2]\;\lraw\;
d\tau^2+\tau^2d\Omega_2^2=dy_1^2+dy_2^2+dy_3^2,\nn \\
&&\star
F_{(3)}\;\lraw\;\frac{\sqrt{3}\vep^{8/3}}{a_0g_s^2Ml_s^2}dx^0\wedge
dx^1\wedge dx^2\wedge dx^2\wedge dy_1\wedge dy_2\wedge dy_3.\\
&&H_{(3)}\;\lraw\;\sqrt{3}\;g_sMl_s^2 dy_1\wedge dy_2\wedge
dy_3\quad \quad\Leftrightarrow\quad
B_{(2)}=\frac{g_sMl_s^2}{2\sqrt{3}}\ep^{ijk}y_i dy_j\wedge dy_k.
\nn
\eeq

To simplify the world-volume action of D3-branes, we set
worldvolume gauge field to zero and take the following (static)
gauge:
\beq{12}\xi_a&=&x_a,\hspace{1.2in}a=0,1,2,3,\nn \\
y_i&=&2\pi l_s\Phi_i, \hspace{1in}i=1,2,3.
\eeq
Then inserting background~(\ref{2}), (\ref{3}) and (\ref{11}) in
non-Abelian action~(\ref{7}) and (\ref{9}), we can expand
lagrangian in powers of $l_s$. The resulted low energy effective
lagrangian as follows$^2$:
\beq{13}
{\cal L}/T_3&=&-\frac{\vep^{8/3}}{a_0g_s^2M^2l_s^4}
\left[p+\frac{4\pi^2l_s^2a_1}{a_0}{\rm Tr}\Phi_i^2
+\frac{16\pi^4l_s^4a_1}{a_0}(\frac{a_1}{a_0}-
\frac{a_2}{a_1}){\rm Tr}(\Phi_i\Phi^i)^2\right] \nn \\
&&-2\pi^2 l_s^2\vep^{4/3}\eta_{ab}{\rm Tr}(\pa^a\Phi_i\pa^b\Phi_i)
+\pi^2\vep^{8/3}{\rm Tr}([\Phi_i,\Phi_j]^2) \nn \\
&&-\frac{2i\pi^2\vep^{8/3}}{\sqrt{3}\;a_0g_sMl_s} \ep_{ijk}{\rm
Tr}(\Phi^i\Phi^j\Phi^k) -\frac{i}{3}4\pi^2g_sl_s [\star
F_{(3)}]_{0123ijk}{\rm Tr}(\Phi^i\Phi^j\Phi^k)+....
\eeq
There are two notable new features in the above lagrangian: 1) The
last two terms in the first line are from warped factor of
background. 2) The first term in the third line is from NS-NS
$B$-field which plays a role in worldvolume action through
eq.~(\ref{7}). It identifies to one from background R-R flux (the
last term in the third line), due to ``no-force'' condition
between D3-brane and imaginary self-dual flux
background\cite{KPV01}.

\subsection{Static solutions}
With the normalization,
$$\Phi_i\;\to\;\frac{\Phi_i}{2\pi l_s\vep^{2/3}\sqrt{T_3}},\hspace{0.5in}
T_3=\frac{1}{g_s(2\pi)^3l_s^4},$$ the effective scalar potential
for the present problem is
\beq{14}
V(\Phi)=V_0+\frac{m^2}{2}{\rm Tr}\Phi_i^2 -\frac{\pi g_s}{2}{\rm
Tr}([\Phi_i,\Phi_j]^2)-\frac{2\pi g_s\lambda}{p^2-1}{\rm
Tr}(\Phi_i\Phi^i)^2+\frac{2if\ep_{ijk}}{3} {\rm
Tr}(\Phi^i\Phi^j\Phi^k),
\eeq
with
\beq{15}
V_0&=&\frac{pT_3\vep^{8/3}}{a_0g_s^2M^2l_s^4}\hspace{0.4in}
m^2=\frac{2a_1\vep^{4/3}}{a_0^2g_s^2M^2l_s^4}, \hspace{0.4in}
f=\frac{\sqrt{6\pi}\vep^{2/3}}{2a_0\sqrt{g_s}Ml_s^2}, \nn \\
\lambda&=&\frac{\kappa^2(p^2-1)}{g_s^2M^2}\hspace{0.6in}
\kappa^2=\frac{4\pi^2a_1}{a_0^2}(\frac{a_2}{a_1}-
\frac{a_1}{a_0})\simeq 1.736\simeq \sqrt{3}.
\eeq
To find the extreme of the potential~(\ref{15}), we obtain the
solution:
\beq{16}
\Phi_i=\frac{m^2}{f}uJ_i,\hspace{0.5in}
u=u_0=\frac{1}{2x}(1\pm\sqrt{1-4x}),
\eeq
where $x$ is defined by
\beq{17}x=\frac{2\pi g_s(1-\lambda)m^2}{f^2}=\frac{8a_1}{3}(1-\lambda),
\eeq
$J_i$ are generators of N-dimensional representation of the
$SU(2)$ group
\beq{18}[J_i, J_j]=i\ep_{ijk}J_k.
\eeq
According to Eq.~(\ref{16}) we have to impose $x\in [0,1/4]$ to
admit non-trivial solution.

\begin{figure}[hptb]
\label{f1} \centering
\includegraphics[width=4in]{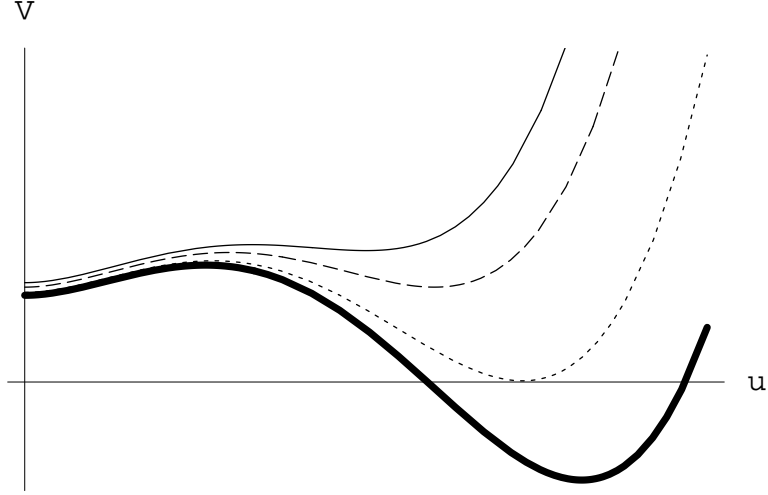}
\begin{minipage}{5in}
\caption{The curve of effective potential $V(u)$ for diverse
values of parameter $x$. The solid line, dash line, dot line and
bold line corresponds to $x=0.24,\;2/9,\;0.2,\;0.18$ respectively,
where local minimum at $u=u_0$ has the properties
$V(u_0)>V_0,\;V(u_0)=V_0,\;V(u_0)=0$ and $V(u_0)<0$ respectively.}
\end{minipage}
\end{figure}

It is useful to rewritten effective potential~(\ref{14}) by new
variable $u$
\beq{19}V(u)=V_0+\frac{p(p^2-1)m^6}{4f^2}
\left\{\frac{1}{2}u^2-\frac{1}{3}u^3+\frac{x}{4}u^4\right\}.
\eeq
In figure 1 we draw the curve of $V(u)$ for several different
values of $x$. We see two solutions presented in Eq.~(\ref{16})
respectively correspond to a local maximum and a local minimum:
\beq{20}
V_{\rm lmax}&=&V_0-\frac{p(p^2-1)m^6}{96x^3f^2}
[1-6x+6x^2-(1-4x)^{3/2}]\geq V_0,\quad\quad {\rm for}\quad\quad
x\in [0,\frac{1}{4}], \nn\\
V_{\rm lmin}&=&V_0-\frac{p(p^2-1)m^6}{96x^3f^2}
[1-6x+6x^2+(1-4x)^{3/2}]\leq V_0, \quad\quad {\rm for}\quad\quad
x\in [0,\frac{2}{9}].
\eeq

One get, therefore, the condition to admit a non-trivial and
stable solution,
\beq{21} 0\leq x\leq
\frac{2}{9}\quad\Leftrightarrow\quad
1-\frac{1}{12a_1}\leq\frac{\kappa^2 (p^2-1)}{g_s^2M^2}\leq 1. \eeq
Because $\kappa$ is a constant of order one, and $1\ll p^2\ll
M^2$, we obtain $g_s\sim p/M\ll 1$. In other words, this solution
is valid in region of weak string coupling, as we expected.
Furthermore, in order to make the expansion in effective
lagrangian~(\ref{13}) be valid, one has
\begin{eqnarray*}
\frac{l_sm^2u}{2\pi l_s\vep^{2/3}\sqrt{T_3}f}\ll 1\quad
\Leftrightarrow\quad
 xg_sM\gg 1\quad \Leftrightarrow\quad  xp\gg 1.
\end{eqnarray*}
This condition in general can be satisfied.

\subsection{Physical explanation}

Geometrically, such the solution represents the fuzzy two-sphere.
The radius of two-sphere can be measured as
\beq{22}R\sim \sqrt{g_sM}l_s^2\left(\frac{{\rm
Tr}\Phi_i^2}{p}\right)^{1/2}\sim \frac{l_sp}{\sqrt{g_sM}}\sim
\sqrt{g_sM}\;l_s.
\eeq
It is just typical scale of background. It is well known that the
local minimum presented in Eq.~(\ref{20}) represents vacuum
corresponding to a spherical D5-brane configuration with topology
$R^3\times S^2$. This spherical D5-brane does not carry net
D5-brane charge. Rather, it would be an ``electric-dipole''-like
configuration due to the separation of oppositely orientation of
$\overline{\rm D3}$-brane on surface of fuzzy $S^2$. From the
viewpoint of dual D5-brane, the process is described by $p$
$\overline{\rm D3}$-branes bounding on D5-brane. At the end of
evolution the $\overline{\rm D3}$-brane dissolve into the U(1)
flux along two-sphere which is just world-volume gauge field
strength on D5-brane. While the pull back of background NS-NS
$B$-field onto world-volume of D5-brane keeps the gauge symmetry
on world-volume. Following steps presented in
\cite{Myers99,KPV01}, it is not hard to prove that the effective
potential~(\ref{19}) can be obtained from world-volume Abelian
action of D5-brane with a constant gauge field strength,
$F_{\theta\phi}=p\sin{\theta}$ with $\theta,\;\phi$ angle
coordinates of $S^2$. We do not repeat the details of calculation
here.

From figure 1 we find two interesting new features on our model:
1) Classically D5-brane can not be nucleated from $\overline{\rm
D3}$-brane if without extra initial conditions, which in general
breaks supersymmetry and makes system be unstable. Quantum
mechanically, however, the vacuum corresponding to $\overline{\rm
D3}$-brane can tunnel to one corresponding to spherical D5-brane.
The tunnelling rate will be computed in next section. 2) From
figure 1 we see that the total energy of system is lowered with
decrease of $x$ and can be even negative. Of course, the negative
energy is unphysical. It just implies that spherical D5-brane is
ripped into D5-$\overline{\rm D5}$ pair by background flux when
size of fuzzy $S^2$ grows to a critical value. The anti-D5 brane
is quickly driven to the apex of KS geometry and annihilate one
unit of background R-R flux. The D5-brane, however, carries
$(K-p)$ D3-brane charge and gets no force from background. Rather,
it receives a small initial velocity from $\overline{\rm D5}$
before $\overline{\rm D5}$-brane disappears and hence slowly moves
to the apex of KS geometry. At the apex it dissolves into $(K-p)$
D3-branes.

\section{Vacuum Tunnelling}

We now turn to compute the nucleation rate of spherical D5-brane
which is generated by quantum tunnelling without extra initial
conditions. It is standard way to find classical solution of
Euclidean action:
\beq{23}
S_{\rm E}=\frac{p(p^2-1)m^4}{4f^2}\int dr\;r^3\{\frac{1}{2}u'^2
+\frac{4f^2V_0}{p(p^2-1)m^4}+m^2(\frac{1}{2}u^2-\frac{1}{3}u^3+
\frac{x}{4}u^4)\},
\eeq
where $u'=du/dr$ with $r$ radial coordinate of $R^4$. The equation
of motion of Euclidean version reads off
\beq{24}u''&=&m^2u(1-u+xu^2), \nn \\ \Rightarrow
  u'^2&=&m^2u^2(1-\frac{2}{3}u+\frac{x}{2}u^2)+c_0.
\eeq
We look for the lowest energy solution whose kinetic energy
vanishes at meta-stable vacuum $V(u=0)$, so that we have $c_0=0$.
The full solution can be obtained explicitly:
\beq{25}u(r)=\frac{3}{1+\cosh(mr)-\sqrt{\frac{9x}{2}}\sinh(mr)}.
\eeq

\begin{figure}[hptb]
\label{f2} \centering
\includegraphics[width=3in]{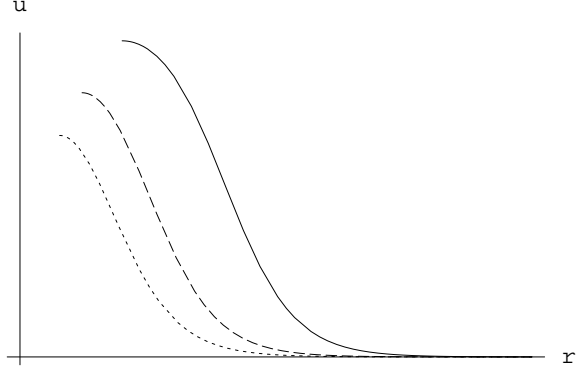}
\begin{minipage}{5in}
\caption{The Euclidean trajectories $u(r)$ for diverse values of
parameter $x$. The solid line, dash line and dot line correspond
to $x=2/9,\;0.2,\;0.15$ respectively, where the trajectories end
with $r=r_*(x)$.}
\end{minipage}
\end{figure}

It looks to be similar to the instanton solution proposed in
\cite{Evslin04}. The tunnelling occurs between $u=0$ and
$u=u_*=\frac{2}{3x}(1-\sqrt{1-\frac{9x}{2}})$. Then the system
classically rolls down to more stable vacuum at
$u=u_0=\frac{1}{2x}(1+\sqrt{1-4x})$. Using the solution~(\ref{25})
we have
\beq{26}u=0\quad&\Leftrightarrow &\quad r\to\infty, \nn\\
u_*=\frac{2}{3x}(1-\sqrt{1-\frac{9x}{2}})\quad&\Leftrightarrow &
\quad r_*=\frac{1}{m}\ln{\frac{1+\sqrt{\frac{9x}{2}}}
{\sqrt{1-\frac{9x}{2}}}}.
\eeq
In figure 2 we plot the Euclidean trajectories $u(r)$ for diverse
values of $x$. It implies that the tunnelling starts from vacuum
at $u=0$ at large $r$. The scalar field stays that meta-stable
vacuum until $r$ close to $r_*$, where it quickly rolls down to
more stable vacuum at $u=u_0$.

Substituting the solution~(\ref{25}) into Euclidean
action~(\ref{23}) one has
\beq{27}S_{\rm E}=\frac{p(p^2-1)m^2}{4f^2}\int_{\td{r}_*}^\infty
d\td{r}\;{\td
r}^3u^2(1-\frac{2}{3}u+\frac{x}{2}u^2)-\frac{V_0}{4m^4}\td{r}_*^4,
\eeq
with $\td{r}=mr$. Using definition on $f,\;m^2$ $V_0$ in
Eq.~(\ref{15}) and (\ref{17}), we obtain that the nucleation rate
of D5-brane is about
\beq{28}\exp{\left(-\frac{a_1}{6\pi g_s}p(p^2-1){\cal
T}(x)\right)}\simeq \exp{\left(-\frac{a_1
M}{8\pi\kappa}p^2\sqrt{1-\frac{3x}{8a_1}}{\cal T}(x)\right)},
\eeq
with
\beq{29}
{\cal T}(x)=\int_{\td{r}_*}^\infty d\td{r}\;{\td
r}^3u^2(1-\frac{2}{3}u+\frac{x}{2}u^2)
-\frac{3a_0^3\kappa^2}{64\pi^2a_1^2(1-3x/(8a_1))}\td{r}_*^4.
\eeq
In figure 3 we show the curve of ${\cal T}(x)$. Hence the
nucleation rate of D5-brane is strong suppressed with the number
of $\overline{D3}$-branes increasing.

\begin{figure}[hptb]
\label{f3} \centering
\includegraphics[width=3in]{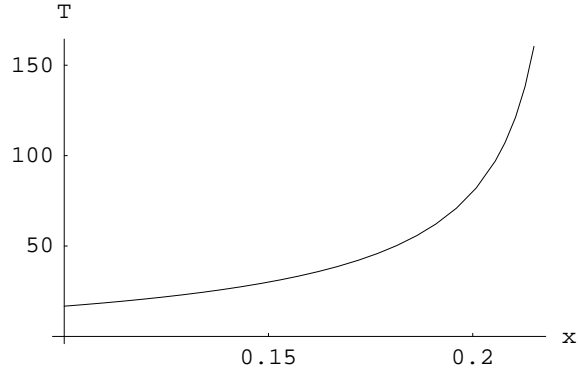}
\begin{minipage}{5in}
\caption{The curve of ${\cal T}(x)$.}
\end{minipage}
\end{figure}

\section{Summary and Discussions}

We have refilled that the dynamics of $p$ $\overline{D3}$-branes
probe the KS geometry in which $M$ units R-R flux and $K$ units
NS-NS flux are presented. An interesting new feature is that
warped factor plays crucial role to world-volume action of
$\overline{D3}$-branes. In other words, it imposes a strong
constraint among $p,\;M$ and $g_s$ in order that the system has
nontrivial static solution. In weak coupling region, i.e.,
$g_s\sim p/M\ll 1$, that nontrivial static solution is a more
stable nonsupersymmetric vacuum which corresponds to D5-brane
``giant graviton'' configuration. Without any extra initial
condition, the nonsupersymmetric vacuum corresponding to
$\overline{D3}$-branes transfers to that spherical D5-brane vacuum
by quantum tunnelling. The tunnelling rate, however, gets strong
suppression with growth of the number of $\overline{D3}$-branes.
The spherical D5-brane can classically be either stable or
unstable. At the end of evolution, the unstable D5-brane
annihilates one unit R-R flux and results in $K-p$ D3-branes.

It is very interesting to compare the evolution of spherical
D5-brane with one of spherical NS-brane considered by KPV
\cite{KPV01}:
\beq{30}
\begin{array}{cccccc}
g_s\ll 1:\quad &\quad p\;\;\overline{\rm D3}\quad & \llraw & {\rm
D5}\;(S^2\;\; {\rm inside}\;\;A{\rm -cycle}) &\llraw &
{\rm stable} \\
&&&& \llraw& (K-p)\;\; {\rm D3} \\
&& {\rm QT} &&  {\rm CE} & \\
g_s\gg 1:\quad &\quad (M-p)\;\; {\rm D3}&\lllaw & {\rm
NS5}\;(S^2\;\; {\rm inside}\;\;B{\rm -cycle})  &\lllaw
& p\;\;\overline{\rm D3} \\
\end{array} \nn
\eeq
where ``QT'' and ``CE'' denote ``quantum tunnelling'' and
``classical evolution'' respectively. Two processes look like to
be inverse each other in brane/flux transmutation and give a good
match under $S$-duality.

Now let us assume that, without any insertion of D3- and $\overline{\rm
D3}$-branes, we have embedded KS geometry into F-theory compactification.
The tadpole condition is satisfied by taking $\chi(X)/24=MK$. After all of
moduli are stabilized, we end with a supersymmetric AdS
vacuum$^3$
\footnotetext[3]{The authors of ref.\cite{RS04} pointed out that there
is no known model that stabilizes the volume modulus with fluxes. We
however still adopt the viewpoint of KKLT model.}
We want to uplift the
AdS vacuum to dS one by adding few $\overline{\rm D3}$-branes while the
topology of CY fourfold $X$ is not changed. It is possible by adding extra
flux simultaneously. For example, we can add one unit R-R flux and $K$
$\overline{\rm D3}$-branes without violation of tadpole condition. Both of
tension of $\overline{\rm D3}$-branes and flux contribute to extra vacuum
energy which uplift the AdS vacuum to dS one. If $g_s\sim K/M\ll 1$,
$\overline{\rm D3}$-branes can tunnel to a classically stable spherical
D5-brane configuration which corresponding to a more stable dS vacuum$^4$.
\footnotetext[4]{From perspective of on-shell string theory, all
dS vacua are meta-stable. They must tunnel to supersymmetric vacua
when scalar fields become large. This region is beyond our
consideration in this present note.} It can also tunnel to a
classically unstable D5-brane configuration which at the end of
evolution annihilate one unit R-R flux. In this case we end with
the original supersymmetric AdS vacuum again.

A more interesting example is to add two unit extra R-R fluxes and
$2K$ $\overline{\rm D3}$-branes into background. When
\beq{31}\frac{(2K)^2}{g_s^2M^2}=\frac{1}{\kappa^2}
(1-\frac{3x_1}{8a_1}), \hspace{1in} x_1<0.2
\eeq
an unstable spherical D5-brane is nucleated by quantum tunnelling
(see Eq.~(\ref{17}) and fig.~1). At the end of evolution it
annihilates one unit R-R flux and remain $K$ $\overline{\rm
D3}$-branes in background. If these residual $\overline{\rm
D3}$-branes can tunnel to a spherical D5-brane again, we require
\beq{32}\frac{K^2}{g_s^2M^2}=\frac{1}{\kappa^2}
(1-\frac{3x_2}{8a_1}), \hspace{1in} x_2\leq 2/9.
\eeq
Combining Eqs.~(\ref{31}) and (\ref{32}) we have
$4x_2-x_1=8a_1\simeq 1.85$, which can not be satisfied. Hence the
residual $\overline{\rm D3}$-branes are stable so that the tension
$K$ $\overline{\rm D3}$-branes and one unit extra R-R flux
contribute extra energy to four-dimensional effective potential.
In other words, it is possible to end with a more stable dS vacuum
in this example.

\acknowledgments{X.-J. Wang thank J.-X. Lu for useful discussion.
This work is partly supported by the NSF of China, Grant No.
10305017, and through USTC ICTS by grants from the Chinese Academy
of Science and a grant from NSFC of China.}

\end{document}